\documentclass[twocolumn]{aastex631}

\usepackage{amsmath,amsmath,ulem,float}
\usepackage{tikz}
\usepackage{ifthen}
\usepgflibrary{fpu}
\graphicspath{{./}{figures/}}

\begin{document}

\author[ 0000-0003-4778-6170 ]{Matthew Belyakov}
\affiliation{Division of Geological and Planetary Sciences, California Institute of Technology, Pasadena, CA 91125, USA}
\author[0000-0002-7094-7908]{Konstantin Batygin}
\affiliation{Division of Geological and Planetary Sciences, California Institute of Technology, Pasadena, CA 91125, USA}

\title{The Perturbation Theory Approach to Stability in the Scattered Disk}

\begin{abstract}
Scattered disk objects (SDOs) are distant minor bodies that orbit the sun on highly eccentric orbits, frequently with perhelia near Neptune's orbit. Gravitational perturbations due to Neptune frequently lead to chaotic dynamics, with the degree of chaotic diffusion set by an object's perihelion distance. \cite{mainpaper} developed a perturbative approach for scattered disk dynamics, finding that, to leading order in semi-major axis ratio, an infinite series of $2:j$ resonances drives the dynamics of the distant scattered disk, with overlaps between resonances driving chaotic motion. In this work we extend this model by taking the spherical harmonic expansion for Neptune's gravitational potential to octupole order and beyond. In continuing the expansion out to smaller semi-major axis limits, we find that the $1:j$ and $3:j$ resonances that emerge in the octupole expansion do not individually set new limits on the stability boundary. Instead, we find that for increasingly Neptune-proximate orbits, resonances of progressively higher index are dominant in explaining the emergence of chaotic behavior. In this picture, the mutual intersections between series of $2:j$, $3:j$, $4:j \dots,$ resonant chains explain local chaotic evolution of SDOs and shape the dynamical distribution of the population at large.
\end{abstract}

\section{Introduction} 
Since the discovery of the first body past Pluto, 1992 QB1 \citep{1992IAUC.5611....1J}, the taxonomy of the Kuiper belt has expanded to include numerous sub-populations, each one shedding light on the evolution of the outer solar system. The distinctions between sub-populations of the Kuiper belt are based on a few criteria: whether or not objects are trapped in a quasi-stable resonance, whether they experience largely secular evolution, or whether they scatter chaotically off of Neptune \citep{2020CeMDA.132...12S}. Within this broad and diverse census of Trans-Neptunian objects (TNOs), a particularly distinctive population is the scattered disk, which is composed of highly eccentric, distant, and relatively planar orbits ($q \gtrsim 30$ au, $a\gtrsim 50$ au, $i\lesssim 20^\circ$). A by-product of the early outwards scattering of planetesimals by Neptune's migration \citep{Gomes2008, annrev}, the scattered disk extends to hundreds of au away from the sun, thus providing a testing ground for dynamical interactions in the outer solar system, as close encounters with Neptune shape the long-term evolution of scattered disk objects (SDOs).  

The scattering dynamics themselves are principally facilitated by the interactions SDOs have with Neptune at perihelion. The strong perturbations from Neptune lead to chaotic evolution as a result of which SDOs are driven out of the solar system, either by diffusion to large enough semi-major axes for galactic effects to ensue, or by becoming centaurs or low-perihelion KBOs (objects with $q<30$, with centaurs also having $a<30$), and finally being ejected by Jupiter. The long-term instability of SDO orbits distinguishes the population from detached Kuiper belt objects, which reside at large enough perihelia to avoid short-period effects from Neptune. Understanding the chaotic dynamics of the scattered disk is an essential component of population studies in the outer solar system, such as those of KBO orbital clustering \citep{P9hypo, 2025arXiv250307146P, 2025ApJ...978..139S}, or analyses based on surface composition \citep{2010AJ....139.1394S, nesvornycolors,  2021AJ....162...19A, 2023PSJ.....4..200P, 2024PSJ.....5..193B}.

Presently, there are three broad approaches for studying the dynamics of the scattered disk. The simplest and most realistic approach is direct integration of test particles (examples include \citealt{2004Icar..172..372F,2008ApJ...687..714V}). Direct integration is useful for understanding population statistics and behavior at large scales across the scattered disk, but does not readily yield insight into how the celestial mechanics of the Sun-Neptune-SDO three body system lead to specific dynamical outcomes. 

The second approach is the mapping approach, in which scattered disk dynamics are treated as a perturbed two-body problem with a periodic kick applied at perihelion. Historically, this approach has relied on the Kepler map to explain the evolution of long-period comets and can be applied to the most eccentric scattered disk objects \citep{1999Icar..141..341M}. Recent work by \cite{2024MNRAS.527.3054H} uses a twist map to describe the evolution of a test particle subject to pericenter-passage kicks. In particular, \cite{2024MNRAS.527.3054H} use this approach to derive resonant widths for mean-motion resonances (MMRs), and by computing a covering fraction of resonances in any given location of phase space are able to predict the onset of chaos in the scattered disk. This mapping approach applies a perturbative treatment to SDOs with long-period orbits, but converges to direct integration for shorter period orbits that are more affected by short-period terms from Neptune. Although the novel method of \cite{2024MNRAS.527.3054H} is able to provide a single curve for the stability boundary of the scattered disk, it leaves open the question of which specific resonant terms are relevant to the dynamical evolution of any given SDO. 

Finally, \cite{mainpaper} characterized the dynamics of the scattered disk using a fully analytic model, seeking to not only place the onset of chaos at a numerical boundary, but also qualitatively explain the underlying machinery of how scattered disk objects go unstable. Using the disturbing function expanded in semi-major axis ratio ($\alpha = a_{\text{N}}/a$) for the circular, restricted, three-body problem, \cite{mainpaper} derived a functional form for the critical value of perihelion ($q_{\text{crit}}$) below which chaos ensues at a given semi-major axis. This result follows from the application of the Chirikov criterion \citep{Chirikov} for resonance overlap to the infinite chain of $2:j$ resonances, which in turn arises from the resonant argument of the quadrupole expansion of the disturbing function\footnote{See equation 5 of \cite{mainpaper}.}. However, this quadrupole analytic approach breaks down at semi-major axis distances below $\sim$400 au, as values of $q_{\text{crit}}$ unphysically drop below 30 au. Therefore, a full analytic understanding of the emergence of chaotic dynamics and scattering in the restricted, circular, three-body problem, remains incomplete in the $a\sim 50-400$ AU region of the scattered disk that most known SDOs occupy. 

In this work, we expand on the approach of \cite{mainpaper}, seeking to construct a predictive theory that does not assume that the Neptune-SDO semi-major axis ratio is a small number. We begin by extending the expansion to the next order -- octupole. This expansion generates two new classes of resonances, $1:j$ and $3:j$, but neither of them individually is able to account for the onset of scattering any better than the quadrupole theory. Quantitatively, this is because each additional term appears to scale in a similar way, meaning that no individual higher order term will define the behavior after the $2:j$ resonant chain begins to break down. Instead, we find that the spacing of resonances of increasing index sets the boundary for chaos in the scattered disk.

This work is organized as follows. In \autoref{sec:oct}, we expand the disturbing function to octupole order and illustrate the inability of individual higher order terms to predict the onset of chaos within 400 au. In \autoref{sec:theory}, we show the results of our general expansion to higher order, and use this expansion to generate a stability boundary that matches orbital integrations of the scattered disk and measurements of the chaos indicator MEGNO. In \autoref{sec:boundcomp}, we arrive at the critical insight that overlap between resonances of all orders at a given point in phase space contribute to chaotic diffusion, and use this notion to develop a modified version of the Chirikov criterion to generate a local boundary for the onset chaotic diffusion. We then combine the local picture of resonance overlap to generate a single stability boundary in \autoref{sec:stabbound}. We summarize our results and look to future work in \autoref{sec:disc}.   

\section{Octupole Theory}
\label{sec:oct}
Expanding the disturbing function out to octupole order largely resembles the quadrupole expansion in \citet{mainpaper}. We begin by stating the assumptions made in the perturbative model for the circular restricted three-body problem, highlighting changes made from previous work. We treat objects in the scattered disk as test particles, however we no longer limit ourselves to the case where the semi-major axis of the SDO is much larger than that of Neptune ($\alpha = a_n/a < 1$ instead of $\alpha \ll 1$). This relaxed assumption necessitates expanding the disturbing function to progressively higher order as $\alpha$ approaches unity. Secondly, we neglect Neptune's eccentricity as the existence of the scattered disk is not contingent on the ellipticity of Neptune's orbit. Finally, we restrict Neptune and the SDO to a common plane, neglecting inclination dynamics.

The disturbing function for the co-planar three-body system with the Sun, Neptune, and an SDO (exterior massless test particle) is given by \citep{1962AJ.....67..300K, Mardling2013}:

\vspace{-0.7cm}
\begin{equation}
\begin{split}
    \mathcal{R} &= \frac{\mathcal{G} m_N}{a} \sum_{\ell=2}^\infty \sum_{m=m_{min}}^\ell \sum_{j'=-\infty}^\infty \sum_{j=-\infty}^\infty c_{\ell m}^2 \mathcal{M}_\ell \\
    &\times \alpha^\ell X_{j'}^{\ell,m}(e_N) X_{j}^{-(\ell+1),m}(e) \\
    &\times \cos(j'M_N - jM + m(\omega_N - \omega)).
\end{split}
\end{equation}

In this series expansion, $\ell$ sets the degree of the expansion, $m$ determines the order of the resonance, and $j$ is the resonant index. In this nomenclature, an $m:j$ resonance is a mean-motion resonance of order $m$ and index $j$, where the SDO completes $m$ orbits in the time Neptune completes $j$ orbits. We expand the above expression to octupole order, setting $\ell_{\text{max}}=3$ (the dynamics of the $\ell=2$ term are explored in \citealt{mainpaper}). The following simplifications and substitutions can be made based on this restriction and the assumptions from above:
\begin{enumerate}
    \item Mass factor $\mathcal{M}_3$ = $\left(m_{\odot}^2-m_N^2\right)/\left(m_{\odot}+m_N\right)^2 \approx 1$.
    \item Hansen coefficients that depend on the eccentricity of Neptune can be neglected: $X_{j'}^{\ell,m}(e_N) = 1$. This also sets $j' = m$, as for $j' \neq m$, $X_{j'}^{\ell,m}(e_N) = 0$.
    \item Set $\omega_N$ to 0 without loss of generality, as $e_\text{N} \to 0$.
\end{enumerate}
From these assumptions, the disturbing function can be simplified to a more tractable form:
\begin{equation}
\begin{split}
    \mathcal{R}^{\text{octupole}} &= \frac{\mathcal{G} m_N \alpha^3}{a} \sum_{m=m_{min}}^3 \sum_{j=-\infty}^\infty c_{3,m}^2 X_{j}^{-4,m} \\ & \times \cos(jM - m(M_N - \omega)),
\end{split}
\end{equation}
where the differences from the quadrupole expansion manifest as an additional factor of $\alpha$, as well as a different set of Hansen coefficients and spherical harmonics.  Substituting possible values of $m$ into $c_{3,m}^2$, we have: \\
$c_{3,m}^2 = \frac{8\pi}{7}(Y_{3,m}(\pi/2, 0))^2 = \begin{cases}
  0 & m = 0, 2 \\
  3/8 & m = 1\\
  5/8 & m = 3\\
\end{cases}$

Eliminating $m=0$ and $m=2$ allows us to isolate the two sets of resonant arguments that will be of interest in the octupole expansion: the $1:j$ and $3:j$ resonances. We can now write the disturbing function as:
\begin{equation}
\label{disturb-oct}
\begin{split}
    \mathcal{R}^{\text{octupole}} &= \frac{\mathcal{G} m_N \alpha^3}{8a} \\&\sum_{j=-\infty}^\infty \left[3 \underbrace{X_{j}^{-4,1}(e) \cos(jM -(M_N-\omega))}_{1:j \text{ resonances}} \right.\\
    & \left.+ 5 \underbrace{X_{j}^{-4,3}(e) \cos(jM - 3(M_N -\omega))}_{3:j \text{ resonances}} \right].\\
\end{split}
\end{equation}

Examining the harmonics, we have $\phi_{1:j} = jM - (\omega - M_N) = j(\lambda-\varpi)-1(\lambda_N-\varpi)$ and $\phi_{3:j} = jM - (\omega - M_N) = j(\lambda-\varpi)-3(\lambda_N-\varpi)$, which are the $1:j$ and $3:j$ MMRs with Neptune respectively. Since the octupole expansion is higher order in the semi-major axis ratio $\alpha$, we would expect the $1:j$ and $3:j$ terms at octupole order to behave differently from the $2:j$ term from the quadrupole expansion. However, as we will show, this is not the case due to the contribution from the Hansen coefficients that arise at octupole order.

The Hansen coefficient at quadrupole order for the $2:j$ resonances is $X_{j}^{-3,2}$, which has been shown to scale linearly with the resonant index $j$ over contours of constant perihelion or as eccentricity approaches unity \citep{2006CosRe..44..160S, 2008CeMDA.100..287S}. $X_{j}^{-4,1}$ and $X_{j}^{-4,3}$ have no known closed form approximations, and combinatorial expansions (see \citealt{1986sfcm.book.....A}) do not offer any direct intuition on the behavior of the coefficient. We instead determine the scaling of $X_{j}^{-4,1}$ and $X_{j}^{-4,3}$ with $j$ numerically by evaluating the coefficient along contours of constant perihelion, $q = a_{\text{SDO}}(1-e) = (j/m)^{2/3} a_{N}(1-e)$. We find that the $j$ dependence of $X_{j}^{-4,1}$ goes as $j^{5/3}$ times a prefactor that depends only on perihelion, holding to less than 1\% for the selected range of perihelion distances. The Hansen coefficient for the $3:j$ resonance $X_{j}^{-4,3}$ also follows the same $j^{5/3}$ scaling, but with a different coefficient than for $X_{j}^{-4,1}$. The coefficients at octupole order therefore pick up an extra factor of $j^{2/3}$ as compared to the linear scaling with $j$ of $X_{j}^{-3,2}$ in the $2:j$ resonance expansion. An analytic demonstration of this scaling of the Hansen coefficients is given in \autoref{sec:scaling}. 

Following the analysis from \cite{mainpaper} and writing down a pendulum-like Hamiltonian for the two resonant terms which follow from the octupole expansion, for which the full derivation can be found in \autoref{sec:1n_res}, we obtain expressions for the Chirikov criterion for the onset of chaotic motion as a function of resonant index, semi-major axis ratio, and the Hansen coefficient for the quadrupole $2:j$ and octupole $1:j$ and $3:j$ terms:

\begin{equation*}
    \begin{split}
        \frac{\Delta a}{\delta a}(2:j) &= \frac{6}{\alpha^{1/2}} \sqrt{\frac{m_n }{M_\odot}X_{j}^{-3,2}}  \\
        \frac{\Delta a}{\delta a}(1:j) &= \sqrt{18 \frac{m_n }{M_\odot}X_{j}^{-4,1}}  \\
        \frac{\Delta a}{\delta a}(3:j) &= \sqrt{270\frac{m_N}{M_\odot} X_{j}^{-4,3}}. \\
    \end{split}
\end{equation*}

\begin{table*}
\begin{tabular*}{\linewidth}{@{\extracolsep{\fill}}c|cccccccc}
    \multicolumn{1}{l}{order ($\ell$)} &&&&&&&& Scaling of $X^{\ell,m}_j$ \\
    \cline{1-1} 
    2 &secular&-&\pmb{$2:j$}&&&&&$\propto j$\\
    3 &-&\pmb{$1:j$}&-&\pmb{$3:j$}&&&&$\propto j^{5/3}$\\
    4 &secular&-&$2:j$&-&\pmb{$4:j$}&&&$\propto j^{7/3}$\\
    5 &-&$1:j$&-&$3:j$&-&\pmb{$5:j$}&&$\propto j^{3}$\\
    6 &secular&-&$2:j$&-&$4:j$&-&\pmb{$6:j$}&$\propto j^{11/3}$\\\hline
    \multicolumn{1}{l}{} &0&1&2&3&4&5&6&$X^{\ell,m}_j\propto j^{(2\ell-1)/3}$\\\cline{2-8}
    \multicolumn{1}{l}{} &\multicolumn{7}{c}{resonant orders $m$, resonant indices $j$}
\end{tabular*}
\caption{Schematic representation of the terms that appear in the expansion of the disturbing function. At each successive order $\ell$ in the expansion, a new resonant index appears. The same resonant index also appears at every other order (even indices appear at even orders), however we only include the term associated with a given resonant index's highest order appearance. Therefore, in all cases except that of the $1:j$ resonance $m=\ell$.}
\label{Tab:pascal}
\end{table*}

Since approximating the Hansen coefficients with a simple functional form becomes more difficult in the octupole expansion than in the quadrupole one, we do not replace the coefficients with a functional form in the expressions for the Chirikov criterion. The expressions for the Chirikov criterion for each resonant chain are similar, with the $1:j$ and $3:j$ expansions picking up a factor of $\alpha^{1/2}$. This factor of $\alpha^{1/2}$ is exactly equal to $j^{-1/3}$, which is canceled out since the Hansen coefficient at octupole order picks up an extra factor of $j^{2/3}$. Remarkably, this scaling will continue for arbitrarily high orders, as at each order, the additional factor in $\alpha^{1/2}$ will be canceled out by the scaling of the Hansen coefficient with $j^{2/3}$. Effectively, each order in the expansion of the disturbing function (assuming high-eccentricity, such that $q_{\text{SDO}}$ is of order $a_N$) is the same up to a constant that does not depend on the resonant order, making the expansion of the disturbing function, in effect, a divergent series. Further discussion of the divergence (and, for low eccentricity, convergence) of the disturbing function can be found in \cite{Ferraz-Mello1994CeMDA}. Instead of working with individual terms (quadrupole, octupole, etc.) in a larger expansion, as is characteristic of expansions involving spherical harmonics, we now proceed to build up our theory by examining the intersections between multiple resonant chains. Just as the intersection between some $2:j$ resonance and its neighboring $2:j+1$ resonance creates chaos, so will the intersection between a $2:j$ resonance and an adjacent $3:j$ resonance.

\section{General Theory for $m:j$ resonances}
\label{sec:theory}

As demonstrated in the previous section, each successive expansion of the disturbing function to higher order yields roughly similar results for the stability boundary each time. Therefore, to improve our resonance-overlap model for the scattered disk, intersection between any members of different resonant classes needs to be accounted for, as intersections between any two resonances will produce a narrow band of chaotic diffusion. To be able to work with arbitrarily high-order resonant chains, we develop a generalization of our theory that can compute the chaotic layer produced by the resonant overlap of any two given $m:j$ and $m':j'$ resonances.  The full derivation is given in \autoref{sec:mj_res}, but we summarize the key points below. For the generic $m:j$ resonant case, the disturbing function is given by:

\begin{equation}
\begin{split}
    \mathcal{R}_{\ell,m} &= \frac{\mathcal{G} m_N \mathcal{M}_\ell \alpha^\ell c_{\ell m}^2 }{a} \sum_{j=-\infty}^\infty  X_{j}^{-(\ell+1),m}\\ &\times\cos(jM - m(M_N-\omega)).
\end{split}
\end{equation}

\begin{figure*}
    \centering
    \includegraphics[width = 0.9\textwidth]{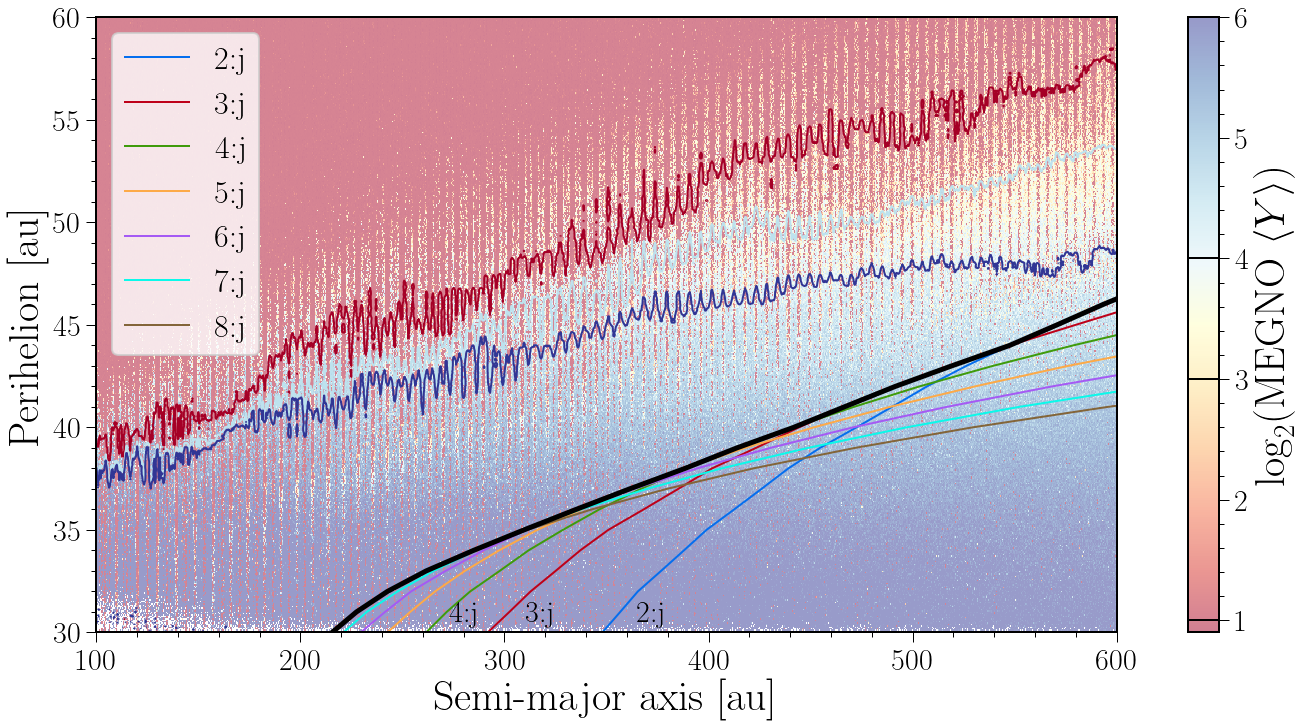}
    \caption{Perihelion stability criteria derived from individual networks of the quadrupole ($2:j$ resonances), octupole ($3:j$ resonances), and successively higher-order ($4\to 8:j$), expansions of the disturbing function. These boundaries do not take into account overlaps between, for example, the 2:7 and 3:10 resonances, but only interactions between same-order resonances. The black line is the result of selecting the best boundary of the individual resonant chains of resonances at each semi-major axis. We overlay these boundaries on the heat map of $\log_2$ of the Mean Exponential Growth of Nearby Orbits (MEGNO) chaos indicator $\langle Y\rangle$ computed in orbital simulations of the Sun, Neptune, and a test particle SDO using \texttt{rebound} \citep{Rein2012A&A}. Three contours roughly show the transition between regular quasi-periodic ($\log_2$(MEGNO)$<1.01$, red) and increasingly chaotic orbits (two contours: $1.01<\log_2$(MEGNO)$<3$, cyan; $4<\log_2$(MEGNO), blue). The lines in the colorbar show the values set by the three contours. While the boundary set by the black line improves upon the single $2:j$ boundary (blue line) determined by \cite{mainpaper} in the 200-400 au region, it still does not match the MEGNO stability map obtained from direct integration. Instead, we must look to the overlap of neighboring resonances of different orders.}
    \label{fig:quadpole_boundary}
\end{figure*}

This full expansion involves three nested infinite series -- iteration over the order of the expansion, the possible resonant orders at that order in the expansion, and all the resonant indices $j$. We show a schematic for the full expansion of the disturbing function in \autoref{Tab:pascal}. At quadrupole order, we only encounter a secular term and the $2:j$ resonant series, at octupole order we encounter the $1:j$ and $3:j$ terms, but at fourth order, we see another set of $2:j$ terms along with the new $4:j$ term, and so on. At each successive order in the expansion of the disturbing function a new resonant term is introduced along with higher-order expansions of previous terms. In our model for resonance overlap, we only select terms where $m=\ell$ (except for the $1:j$ resonances where $\ell=3$ but $m=1$) which is the first appearance of any given resonant term. These higher-degree terms, such as $\ell=4, m=2$ in \autoref{Tab:pascal} which also describes the behavior of $2:j$ resonances, will contribute to chaotic diffusion around the separatrices of resonances, as described in \cite{Morbidelli1997PhyD}. However, the inclusion of higher-degree terms will not significantly widen associated resonances, and therefore can be neglected in our analysis.

\begin{figure*}
    \centering
    \includegraphics[width=0.85\textwidth]{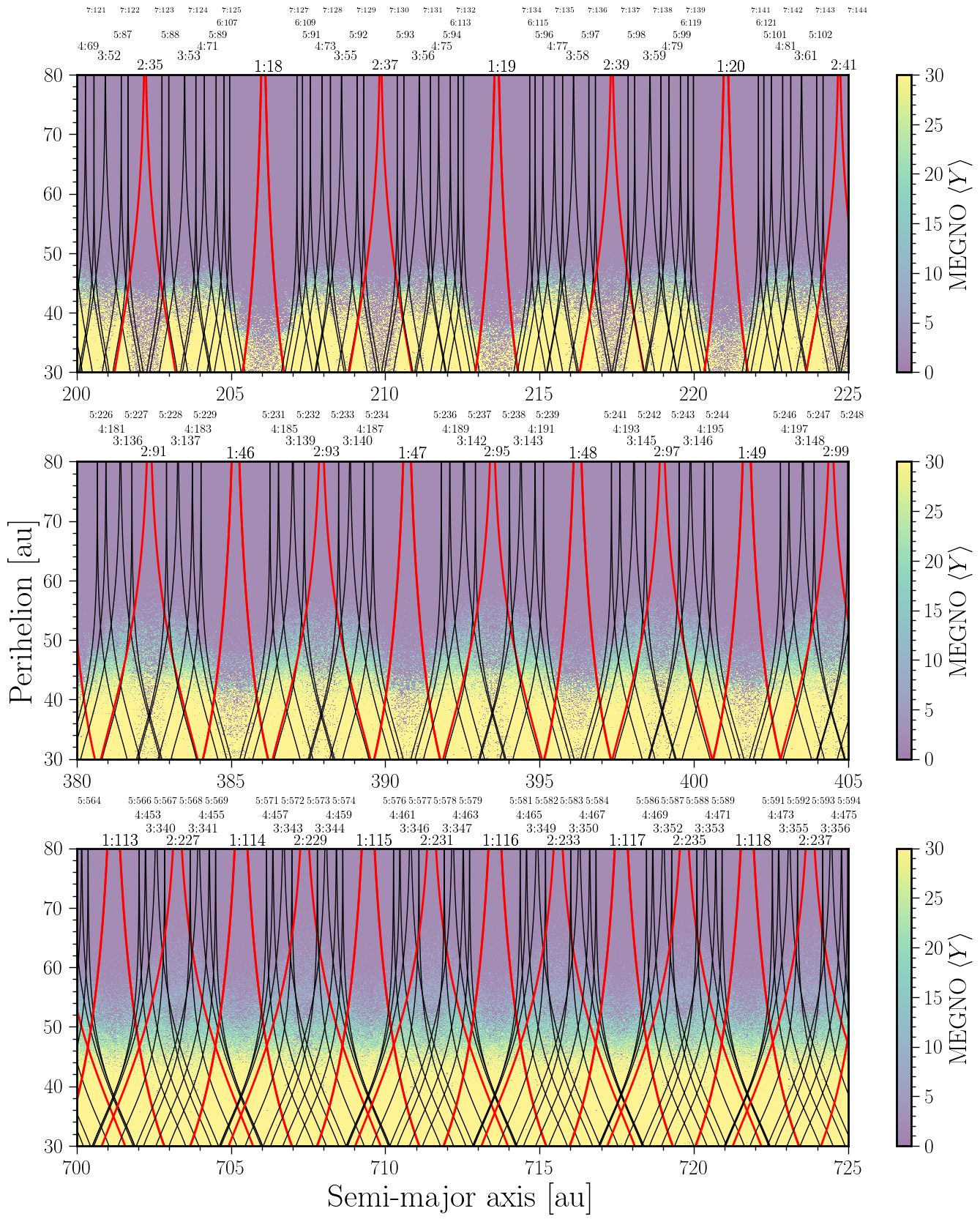}
    \caption{Three heat maps of the chaos indicator MEGNO $\langle$Y$\rangle$ on a semi-major axis versus perihelion plane, derived from simulations of the Sun, Neptune, and test particle scattered disk objects. Purple regions correspond to initial conditions that lead to regular motion, whereas particles initialized in yellow regions evolve chaotically. The nominal locations and widths of mean motion resonances are shown with red ($1:j$, $2:j$) and black ($3:j$ and higher) lines and are labeled above the heatmaps.}
    \label{fig:bigplot1}
\end{figure*}

As we show in \autoref{sec:mj_res}, the width of a given $m:\chi$ resonance is given by
\begin{equation}
\label{eq:reswidth}
\begin{split}
    \Delta a_{m:\chi} = \frac{8a}{\sqrt{3}} \sqrt{\frac{m_N}{M_\odot}} \cdot \sqrt{\alpha^m \mathcal{M}_m c_{m,m}^2 X_{\chi}^{-(m+1),m}}.
\end{split}
\end{equation}
Equipped with the expression for the width of any resonance to arbitrary order and the distance between any two given resonances, we can now analytically determine whether any two resonances overlap for a given eccentricity/perihelion value. The Hansen coefficient is the only component of our expression that requires a numerical calculation, which we evaluate by numerical integration (see equation B19 in Appendix B4 of \citealt{Mardling2013}). 

Arguably the most naive approach to improving on the quadrupole boundary from \cite{mainpaper}, we can simply find the resonant widths of successively higher index resonance chains and compute the stability boundary for each chain. Then, at every increment in semi-major axis we can select the chain that has resonance overlap at the highest perihelion, and use that as our boundary. In \autoref{fig:quadpole_boundary}, we show the result of generating stability boundaries from overlap of resonant chains from $2:j \dotsc 8:j$, and then selecting at each semi-major axis the highest perihelion value out of the individual boundaries generated. There is clear improvement in the 200-400 au range, with the boundary taking on a more linear functional form, monotonically decreasing in perihelion below 600 au. However, as shown by direct integration of the three-body problem with the chaos indicator MEGNO (mean exponential growth factor of nearby orbits; \citealt{Cincotta2000A&AS})\footnote{We describe these simulations and the MEGNO criterion, which are used throughout the paper, in \autoref{sec:simulations}.}, the part of phase space in which chaotic diffusion is occurring extends to much higher perihelia than the boundary set by the overlap of individual resonant chains of the same order $\ell$ would suggest. As discussed in the previous section, we do not expect these successive expansions to arbitrary order in the disturbing function to result in a significant improvement -- the Hansen coefficient scales similarly at all orders of the expansion of the disturbing function.

\subsection{Overlap between different resonant orders}
\label{sec:boundcomp}

The more precise way of computing the stability boundary from resonance overlap is to include the overlap between all resonances near a given location in phase space. In developing their criterion for the onset of chaos through resonance overlap, \cite{2024MNRAS.527.3054H} determine a local density of resonances based on work from \cite{Quillen2011MNRAS} and \cite{Hadden2018AJ}. Between any two given $1:j,1:j+1$ resonances, \cite{2024MNRAS.527.3054H} evaluate a ``covering fraction of resonances'' such that if the value is greater than or equal to one, overlaps between resonances cover the entire area between the adjacent $1:j$ resonances, thus chaotic diffusion ensues. This approach sets an upper bound on any possible chaotic diffusion in the three-body problem, but obscures the fine resonant structure that governs the dynamical evolution of the scattered disk.

\begin{figure}[h!tb]
    \includegraphics[width=0.48\textwidth]{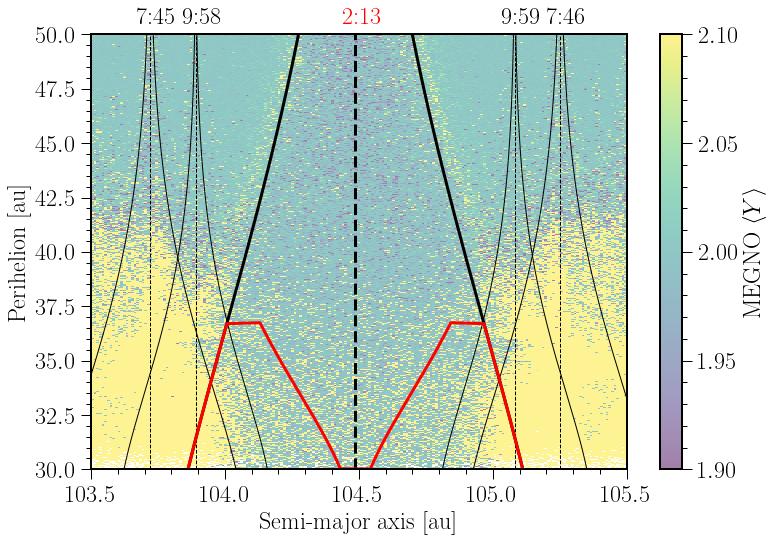}
    \includegraphics[width=0.48\textwidth]{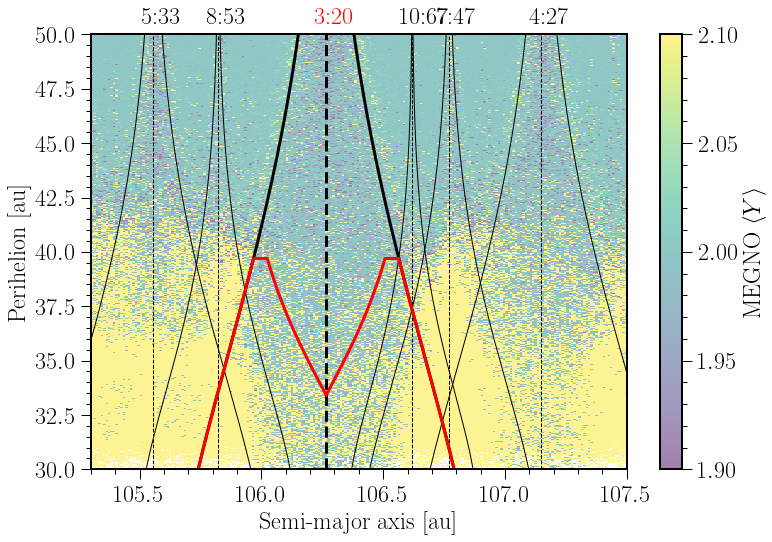}
    \caption{Top and bottom panels demonstrate graphically our modified Chirikov criterion (shown in red) based on the relative strength of resonances (individual resonances shown in black lines). In the top panel, the strong $2:13$ resonance overlaps only slightly with the much weaker $7:45$, $9:58$, $9:59$, and $7:46$ resonances. The modified stability criterion identifies that trajectories in the middle of the $2:13$ are more stable than those closer to the neighboring higher-order resonances. For the $3:20$ resonance in the bottom panel, the more stable area in the center of the resonance is smaller, as the ratio of the widths of nearby resonances is closer to unity. These individual boundaries obtained for every resonance out to arbitrary order can then be combined into a single boundary for stability in the Scattered disk.}
    \label{fig:explain}
\end{figure}

More specifically, the covering fraction approach does not account for the spacing between resonances. Resonances appear at integer ratios -- rational numbers -- consequently the relative spacing and local density between a pair of $1:j,1:j+1$ resonances is given by the Farey sequence \citep{PhysRevSTAB.17.014001, Hadden2018AJ}. The Farey sequence is the sequence of completely reduced fractions, e.g. $F_{4} = \{\frac{0}{1}, \frac{1}{4}, \frac{1}{3}, \frac{1}{2}, \frac{2}{3}, \frac{3}{4}, \frac{1}{1}\}$. For sequences F$_{n}$ with small $n$, there tend to be more terms that are closer to $1/2$ than to $0/1$ or $1/1$. In the first 20 sequences $F_{1}\dotsc F_{21}$, all sequences have the same number or more terms closer to $1/2$ than to $0/1$. Given our previous results on the widths of $1:j$ and $2:j$ resonances in \autoref{sec:oct}, we can hypothesize that the increased density of resonances around the $2:j$ resonances should cause an earlier onset of chaos with increasing eccentricity than in the neighborhood of $1:j$ resonances. 

In \autoref{fig:bigplot1} we show three panels with the ``forest'' of resonances from $1:j$ out to $7:j$ around 200, 400, and 700 au in semi-major axis, overlaid on top of heat maps of the MEGNO chaos indicator. At $a\sim200$ au, particles in the vicinities of both the $2:j$ and $1:j$ resonances tend to be less chaotic than between these resonances. At $a\sim400$ au, individual $2:j$ resonances begin to overlap as shown in \cite{mainpaper} and the chaotic layer starts to become more smooth, though $1:j$ resonances still generate barriers that can locally slow down diffusive transport. Finally, at $a\sim700$ au, resonances become so dense that the overlap between $2:j$ resonances becomes a good approximation for the onset of chaos.

Given the structure of resonances shown in \autoref{fig:bigplot1}, an improved boundary for stability based on resonance overlap needs to incorporate the onset and extent of resonance overlap with respect to perihelion and semi-major axis. In the typical formulation of the Chirikov criterion, if the sum of the half-widths of two adjacent resonances divided by the distance between their nominal locations is greater than unity, then motion of particles within their domains will be chaotic. This can be expressed as: 
\begin{equation}
    \text{chaos ensues when  } \frac{\Delta a - \Delta a_{\textrm{adjacent}}}{[a]-[a]_{\textrm{adjacent}}} > 1,
\end{equation} where $\Delta a$ is the resonance width and $[a]$ is the nominal location of the resonance. However, we are examining resonances of different widths, whereas in the Standard map resonance widths are assumed to be equal \citep{Chirikov}. For example, if a strong $2:j$ resonance is just barely touching a much weaker $7:j$ resonance, we do not expect trajectories close to the nominal center of the $2:j$ resonance to be chaotic or strongly perturbed. However, as described in Section 4 of \cite{Chirikov}, adjacent resonances will deform each others separatrices, and resonances can be everywhere dense. We therefore modify the Chirikov criterion to account for the density of resonances and order of magnitude differences in resonance strength. 

Let us take two resonances with widths $\Delta a_1(q)$ and $\Delta a_2(q)$ that are functions of perihelion, and nominal centers $[a]_1$ and $[a]_2$ such that $[a]_2>[a]_1$. We first determine the perihelion $q_{\text{crit}}$ for which the usual Chirikov criterion is met. For perihelia below $q_{\text{crit}}$, we determine the extent of resonance overlap by computing $\Delta a_1(q) + [a]_1$, the width of one resonance added to its nominal center, and evaluating whether it reaches the center of the adjacent resonance $[a]_2$. For values of $q$ where $\Delta a_1(q) + [a]_1 \geq [a]_2$, all trajectories with $a \in [a]_2\pm\Delta a_2(q)/2$ are chaotic, while for values of $q$ where $\Delta a_1(q) + [a]_1 < [a]_2$, trajectories within the region $a_1(q) + [a]_1 < a < 2[a]_2 - (a_1(q) + [a]_1)$ are stable, but are unstable for all other $a \in [a]_2\pm\Delta a_2(q)/2$. In plain language, when a weaker resonance overlaps a stronger one, we only allow the weaker resonance to interact with the stronger resonance up to the full-width extent of the weaker resonance. If the full width of the smaller resonance does not reach the vicinity of the nominal center of the stronger resonance, then particles in that vicinity fall outside of our boundary for chaos, reproducing the behavior of particles deep in the $1:j$ and $2:j$ resonances shown in the first and second panels of \autoref{fig:bigplot1}. We show the process of computing our modified Chirikov criterion in \autoref{fig:explain}, with two examples of strong $2:j$ and $3:j$ resonances overlapping nearby weaker resonances producing a boundary that is at lower perihelion towards the middle of the wider resonance. Trajectories that begin close to the nominal center of the $2:13$ resonance in the left panel of \autoref{fig:explain} are stable, while trajectories that begin closer to the separatrix will chaotically diffuse out.  

\subsection{The Stability Boundary}
\label{sec:stabbound}

\begin{figure*}
    \centering
    \includegraphics[width=0.9\textwidth]{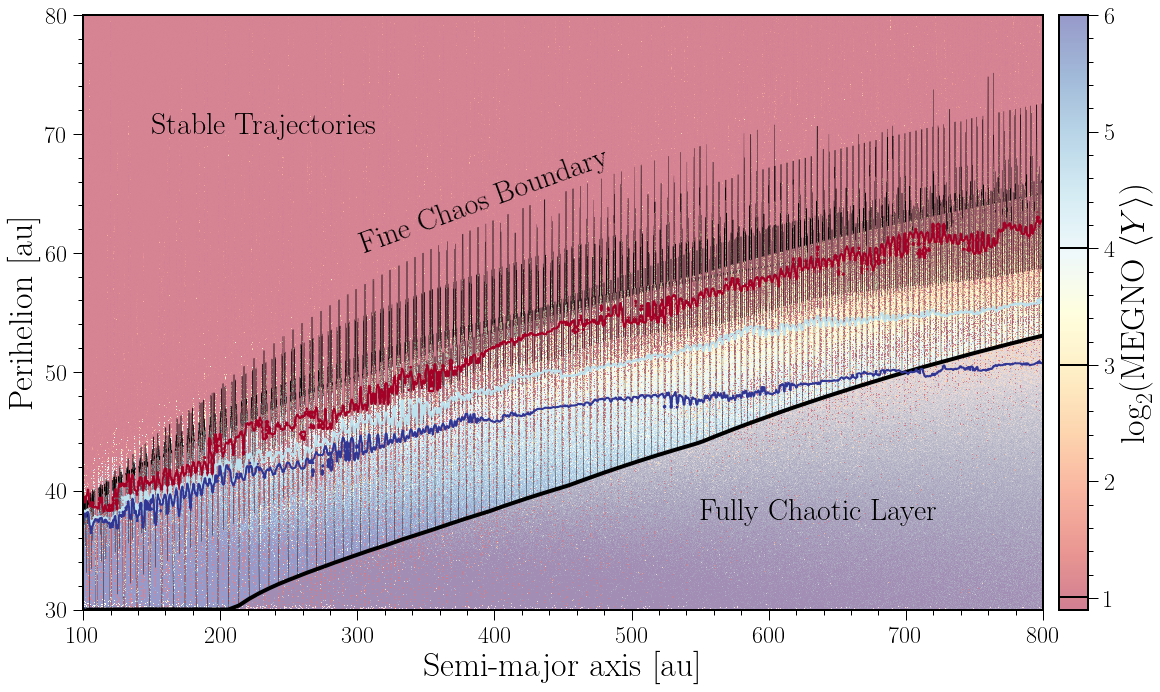}
    \caption{Our full stability boundary plotted on top of the MEGNO map for the circular restricted three-body problem. Due to the inclusion of resonances from $1:j$ to $10:j$, we see extremely fine comb-like structure appear due to the relative placement and strength of resonances. Selecting the minimum values of the chaos boundary, we get a secondary boundary below which there do not appear to be significant zones of stability.}
    \label{fig:megnodiagram}
\end{figure*}

To derive our stability boundary, we begin by computing the local stability boundary at resonances of orders $1:j$ to $10:j$ in the 100-1000 au range using our modified Chirikov criterion as demonstrated in \autoref{fig:explain}. We then need to combine these locally computed boundaries into a single boundary. The first approach we can take is to select the maximum perihelion value at each semi-major axis from each resonant order, essentially tracing over the maxima of the local boundaries to obtain a single coherent boundary. This will correspond to the overall boundary for where chaotic diffusion begins in the scattered disk, and should be roughly equivalent to the semi-analytic result of \cite{2024MNRAS.527.3054H}. We show this boundary in \autoref{fig:megnodiagram} plotted on top of the MEGNO map from previous figures. The resulting boundary is excellent at characterizing locally the behavior of particles in the circular restricted three-body problem out to 100 au, and predicts that at $\sim$200 au, a secondary layer begins to form (shown by the thick black line), within which there are no long-term stable trajectories. This roughly corresponds to the boundary in \autoref{fig:bigplot1}, inside which resonance overlap becomes dense enough that even near the centers of nominal resonances there is no stabilizing effect. The fine comb of resonances shown in \autoref{fig:megnodiagram} creates a transition between the entirely chaotic region of phase space at very high eccentricity and relatively stable orbits at lower eccentricity. In this transitionary region, we should expect to SDOs locked in resonances, or, alternatively, on chaotic orbits that have diffusion slowed by the interactions with strong resonances.

\begin{figure*}
    \centering
    \includegraphics[width=0.9\textwidth]{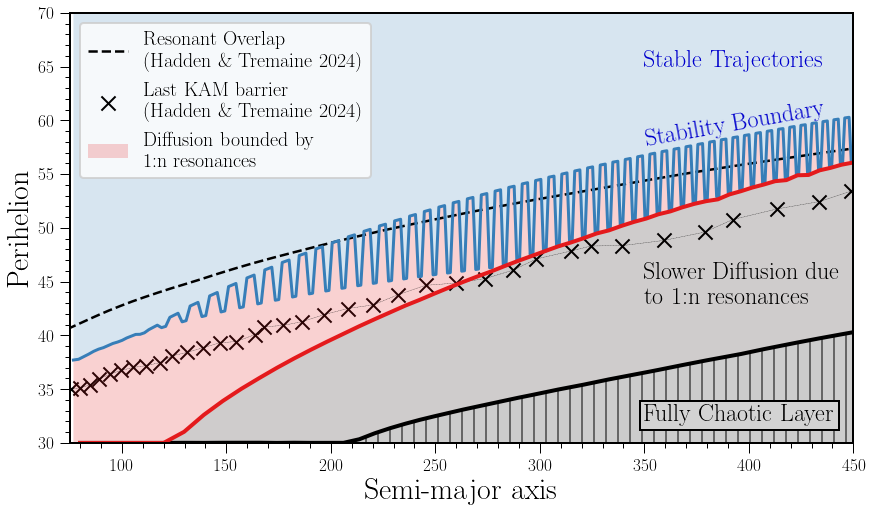}
    \caption{Schematic for the different scattering regimes in the circular restricted three-body problem as derived from resonance overlap. In the black dashed line is the resonant overlap criterion from \cite{2024MNRAS.527.3054H}, and the black crosses are the boundary for which motion is bounded by KAM tori again from \cite{2024MNRAS.527.3054H}. Our stability boundary in blue lies between these two boundaries, and shows that with increasing semi-major axis, stability as a function of $(q,a)$ has significant local dependence. Below the blue line, trajectories diffuse more rapidly with lower perihelion, with $1:j$ resonances serving as barriers to semi-major axis evolution. The red line is the maximum value for the overlap criterion computed at $1:j$ resonances. Below this boundary, the black line and the barred region show the part of phase space where $1:j$ resonances no longer act as barriers to diffusion, as the collection of resonances near them grow strong enough to have strong overlap.}
    \label{fig:diagram}
\end{figure*}

In order to better understand the fine chaos boundary and the transition between the fully chaotic layer and stable region in \autoref{fig:megnodiagram}, we can leverage that each local boundary is associated with a specific resonant order in order to then obtain multiple boundaries to more granularly characterize the dynamical behavior of SDOs. For example, the boundaries in \autoref{fig:explain} can be converted to a single point at the center of the nominal resonance in semi-major axis and a perihelion value equal to either the maximum of the local boundary or the perihelion value at the center of the nominal resonance, thus generating a single boundary. These two options for selecting the perihelion of the boundary are physically meaningful. For example, a global boundary set using the value at the center of the nominal resonance of the local boundaries associated with the $1:j$ resonances is the boundary below which no trajectory can be stabilized by librating around a $1:j$ resonance. Alternatively, the boundary obtained from the maxima of the local $1:j$ boundaries is the line above which SDOs are bounded in semi-major axis between adjacent $1:j$ resonances, and will not pass outside of the adjacent $1:j$ resonances unless they are transported along the separatrices of the $1:j$ resonances, significantly limiting their diffusion. 

In \autoref{fig:diagram} we show our schematic for the different regimes particles will experience in the scattered disk, plotting these against the boundaries derived in \cite{2024MNRAS.527.3054H}. Our stability boundary shown in blue is defined as the perihelion of the local boundary for $10:j$ resonances at their nominal centers. This boundary closely follows the resonant overlap boundary from \cite{2024MNRAS.527.3054H}, but falls off at lower perihelia. Below this line and above the red line defined by the maximum value for the overlap criterion at $1:j$ resonances is the zone in which diffusion is severely impeded at the locations of $1:j$ resonances. A similar idea was suggested in Section 4.4 of \cite{2024MNRAS.527.3054H}, who find that particles in the vicinity of $1:j$ resonances are ``sticky'' and impede diffusion. Below the peak of the local $1:j$ boundaries is the most chaotic region, which we split into two parts. One is the fully chaotic layer that matches the initial result in \autoref{fig:bigplot1} or the lower boundary in \autoref{fig:megnodiagram}, and above it is a layer where we expect slower diffusion from the stickiness of $1:j$ resonances as particles that begin close to the centers of the $1:j$ resonances can remain on librating orbits. We provide approximate functional fits to these boundaries in \autoref{sec:appxfits}. We also make the boundaries shown in \autoref{fig:diagram} and \autoref{fig:megnodiagram} available online\footnote{\url{https://doi.org/10.22002/men4e-m4a20}} \citep{belyakov_batygin_2025}.

\section{Discussion}
\label{sec:disc}

With approximately 10$^4$ scattered disk objects predicted to be discovered by Vera Rubin Observatory in the coming decade \citep{Kurlander2025AJ}, the striking complexity of this population will invigorate studies both of the orbital evolution of individual objects and the scattered disk as a whole. Previous theoretical characterizations of TNO scattering leveraged the detection of individual objects sticking to $1:j$ mean-motion resonances with Neptune to argue for the importance of $1:j$ resonances in sculpting the scattered disk \citep{2004AJ....128.1418P, 2007Icar..192..238L, 2019CeMDA.131...39L, 2024PSJ.....5..135G}. These works, however, do not characterize the behavior of long-period SDOs, which Vera Rubin Observatory will excel at detecting. The approach of \cite{2024MNRAS.527.3054H} determined the onset of chaos in the scattered disk using the mapping approach, providing a clear boundary in perihelion as a function of semi-major axis, below which SDO orbits chaotically diffusive. 

The goal of our perturbative treatment of the circular restricted three-body problem has been to uncover the machinery of the network of resonances that underlie the scattered disk. We expand on the theory of \cite{mainpaper}, which leveraged the quadrupole expansion of the disturbing function \citep{1962AJ.....67..300K, Mardling2013} to determine that the overlap between $2:j$ resonances drives chaotic evolution in SDOs past 400 au. By taking the expansion of the disturbing function to octupole order and repeating the exercise of \cite{mainpaper} for $1:j$ and $3:j$ resonances, we find that in isolation, each order in the expansion of the disturbing functions yields a similar stability boundary that does not significantly improve on the one derived from $2:j$ resonance overlap alone. Using the intuition obtained from the octupole expansion, we generalize our perturbative treatment of the three-body problem by determining the resonance widths and overlaps between mean-motion resonances out to arbitrary order. We find that chaotic transport is facilitated by the density of resonances, such that for lower semi-major axes, it is the mutual overlap between harmonics of ever-higher order that results in diffusive behavior. By modifying the Chirikov criterion to account for the lesser amount of diffusion from the overlap of strong and much weaker resonances, we are able to generate a stability boundary from the overlap of all $1:j$ to $10:j$ mean-motion resonances with Neptune as shown in \autoref{fig:megnodiagram}. 

Our key result is the mapping of the geography of resonances that facilitate scattering (\autoref{fig:diagram}). Past 400 au, a single network of resonances -- the $2:j$ as shown by \cite{mainpaper} -- are sufficient to explain the onset of chaos in the scattered disk. Closer in, the Farey sequence banded resonance structure drives diffusion in two distinct diffusive regimes. One is a zone of complete overlap in a fully chaotic layer, where no resonance provides a barrier to transport. The second is a zone of slower diffusion, in which $1:j$ resonances emerge as increasingly strong barriers to diffusion as eccentricity is lowered. This regime in which $1:j$ resonances act as a boundary for chaotic evolution, or in other words, are sticky, is a direct consequence of the Farey sequence being the representation of orbital commensurabilities. 

Future work will address how the stability boundary for the scattered disk is altered by the introduction of inclination resonances. Scattered disk objects acquire their inclinations either through close encounters with Neptune, or by exchanging eccentricity for inclination through the von Zeipel-Lidov-Kozai mechanism when trapped in mean-motion resonances \citep{Gomes2008}. The inclination of the scattered disk has typically been characterized as low (90\% of SDOs are below 40$^\circ$), with the persistent caveat that observational bias favors detection of objects in the ecliptic plane. Recent efforts by \cite{Bernardinelli2022ApJS} in discovering Kuiper belt objects at high-ecliptic latitudes have not significantly changed the observed inclination distribution of SDOs. We anticipate that the addition of inclination-dependent terms in the disturbing function will cause diffusion to diminish at high eccentricity, in line with predictions by \cite{Gallardo2019Icar}.

\bibliography{references}{}
\bibliographystyle{aasjournal}

\appendix

\section{Scaling of Hansen coefficients}
\label{sec:scaling}
\begin{figure*}[h]
\centering
    \includegraphics[width=0.6\textwidth]{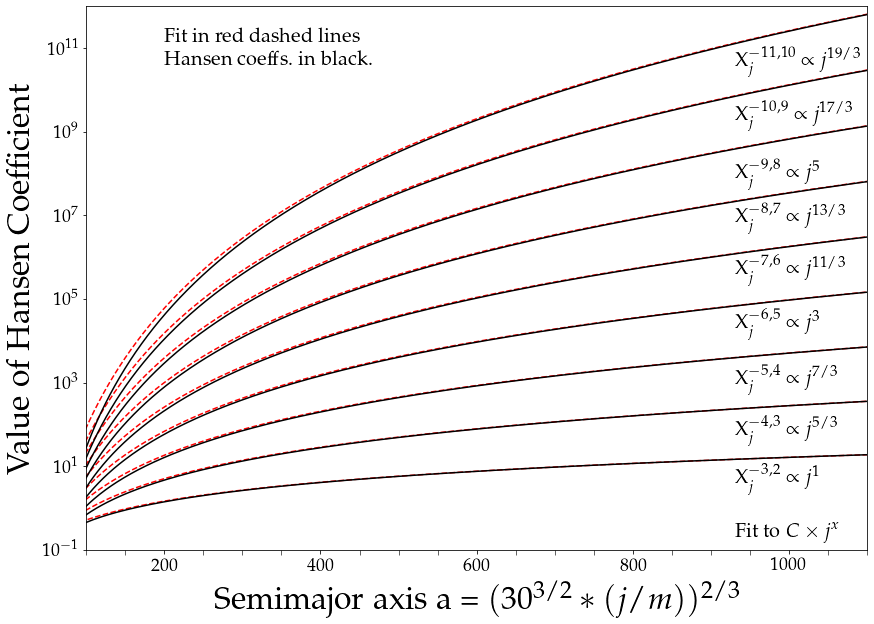}
    \caption{Numerically computed Hansen coefficients $X^{\ell,m}_j$ at $q=40 au$ in black, plotted at the nominal semi-major axis of each $m:j$ resonance.}
    \label{fig:hansen}
\end{figure*}

We begin with the recurrence relation for the Hansen coefficients from \cite{1986sfcm.book.....A}.
\begin{equation}
\begin{split}
    (1-e^2) X^{n, m}_j &= X^{n+1,m}_j + \frac{e}{2}\left(X^{n+1, m+1}_j + X^{n+1, m-1}_j\right) \\
    X^{n, m}_j &= \frac{1}{(1-e^2)}X^{n+1,m}_j + \frac{e}{2(1-e^2)}\left(X^{n+1, m+1}_j + X^{n+1, m-1}_j\right)
\end{split}
\end{equation}
Using $e=1-q/a=1-q/(a_N \cdot j^{2/3})$, we can rewrite the above equation as:
\begin{equation}
\begin{split}
    X^{n, m}_j = \frac{a_N^2 \cdot j^{4/3}}{2 q \cdot a_N \cdot j^{2/3} - q^2}X^{n+1,m}_j + \frac{1}{2} \frac{a_N^2 \cdot j^{4/3} - q \cdot a_N \cdot j^{2/3}}{2 q \cdot a_N \cdot j^{2/3} - q^2}\left(X^{n+1, m+1}_j + X^{n+1, m-1}_j\right)
\end{split}
\end{equation}
In the limit where $q$ is not much larger than $a_N$ (large $j$), this expression approaches:
\begin{equation}
\begin{split}
    X^{n, m}_j = \frac{j^{2/3}}{2}X^{n+1,m}_j + \frac{j^{2/3}}{2}\left(X^{n, m+1}_j + X^{n, m-1}_j\right).
\end{split}
\end{equation}
Given that $j$ is negative in our analysis for an internal perturber, we have recovered the $j^{2/3}$ scaling of the Hansen coefficient with increasing degree of the expansion. We also show this scaling numerically in \autoref{fig:hansen}. For small $j$, the fit becomes poor for Hansen coefficients associated with very large $\ell$, however it holds extremely well for large $j$ at all orders.

\section{Derivation of resonance width for $1:j$ and $3:j$ resonances} \label{sec:1n_res}
Starting with the octupole order disturbing function in Equation \eqref{disturb-oct}, we can write down two model Hamiltonians for an SDO interacting with the $1:\chi$ and $3:\chi$ resonances respectively. We only give the derivation for the special case of the resonance width of a $1:\chi$ resonance while the widths of $3:\chi$ resonances can be evaluated from the general case for $m:j$ resonant widths, given in the section B of the appendix. The Hamiltonian for the $1:j$ resonances using the pendulum model for resonances can be written as:

\begin{equation}
    \mathcal{H} = -\frac{\mathcal{G}M_\odot}{2a} - \frac{3\mathcal{G} m_N \mathcal{M}_3 \alpha^3}{8a} \sum_{j=-\infty}^\infty  X_{j}^{-4,1}(e) \cos(jM - (M_N - \omega)) + \mathcal{T}
\end{equation}

We can use Delaunay Variables as given in ch. 2 of Murray and Dermott to rewrite the Keplerian term as:
\begin{equation}
    -\frac{\mathcal{G}M_\odot}{2a} = -\frac{1}{2}\left(\frac{\mathcal{G}M_\odot}{L} \right)^2
\end{equation}

This can be Taylor expanded around the nominal action $[L] = \sqrt{\mathcal{G} M_\odot a_N (\chi/1)^{2/3}}$:
\begin{equation}
\begin{split}
    -\frac{1}{2}\left(\frac{\mathcal{G}M_\odot}{{L}}\right)^2  = -\frac{1}{2}\left(\frac{\mathcal{G}M_\odot}{{[L]}}\right)^2 
     + \underbrace{\left(\frac{\mathcal{G}M_\odot}{{L}}\right)^2\left(\frac{\delta {L}}{[{L}]}\right) - \frac{3}{2}\left(\frac{\mathcal{G}M_\odot}{{L}}\right)^2\left(\frac{\delta {L}}{[{L}]}\right)^2}_{[n]\left(\delta L-\frac{3}{2}\frac{\delta{L}^2}{{[L]}}\right)}
\end{split}
\end{equation}
We define a type-2 generating function:
\begin{equation}
    \mathcal{F}_2 = \underbrace{(\chi l - (n_N t - g))}_{\phi}\Phi + \underbrace{(l)}_{\psi}\Psi + \underbrace{(t)}_{\xi} \Xi
\end{equation}
with transformation equations between the old and new actions defined as:
\begin{equation}
    \begin{split}
        &\delta L = \frac{\partial \mathcal{F}_2}{\partial l} = \chi \Phi + \Psi\\
        &\delta G = \frac{\partial \mathcal{F}_2}{\partial g} = \Phi\\
        &\delta \mathcal{T} = \frac{\partial \mathcal{F}_2}{\partial t} = \Xi - n_N \Phi
    \end{split}
\end{equation}
We apply these transformations to our Hamiltonian, now isolating just one resonance $1: \chi$:
\begin{equation}
        \mathcal{H}_\chi = - \frac{3\mathcal{G} m_N \mathcal{M}_3 \alpha^3}{8a}  X_{\chi}^{-4,1} \cos(\phi) + \Xi - n_N \Phi + [n]\left(\chi \Phi + \Psi -\frac{3}{2}\frac{(\chi \Phi + \Psi)^2}{{[L]}}\right) -\frac{1}{2}\left(\frac{\mathcal{G}M_\odot}{{[L]}}\right)^2
\end{equation}
Simplifying and removing constant terms (those with only $\Psi$):
\begin{equation}
        \mathcal{H}_\chi = - \frac{3\mathcal{G} m_N \mathcal{M}_3 \alpha^3}{8a}  X_{\chi}^{-4,1} \cos(\phi)  - \frac{3n_N \chi}{2[L]}\Phi^2 - \frac{3n_N }{[L]}\Psi\Phi
\end{equation}
We can further simplify this by completing the square using the linear action term:
\begin{equation}
    \begin{split}
        \mathcal{H}_\chi &= - \frac{3\mathcal{G} m_N \mathcal{M}_3 \alpha^3}{8a}  X_{\chi}^{-4,1} \cos(\phi)  - \frac{3n_N\chi}{2[L]}\underbrace{\left(\Phi + \frac{\Psi}{\chi}\right)^2}_{\Tilde{\Phi}^2} 
    \end{split}
\end{equation}
where $\Tilde{\Phi}$ is the new action conjugate to the angle $\phi$. The resonance width of a pendulum is given by $\Delta\Tilde{\Phi} = 4 \sqrt{\gamma/\beta}$, where, $\Delta \delta L = \chi \Delta \Tilde{\Phi}$. However, we want the width, $\Delta a$, which we can find using $\Delta \delta L = \Delta (L-[L]) = \Delta L$, where $[L]$ is constant, such that $\Delta[L]= 0$, which gives us 
\begin{equation}
\begin{split}
    \Delta \delta L &= \Delta(\sqrt{\mathcal{G}M_\odot a}) = \frac{\Delta a}{\sqrt{a}}\frac{\sqrt{\mathcal{G}M_\odot}}{2}\\
    \Delta a &= \frac{2\sqrt{a} \Delta \delta L}{\sqrt{\mathcal{G}M_\odot}} = \frac{8 \chi \sqrt{a \cdot \gamma/\beta}}{\sqrt{\mathcal{G}M_\odot}}.
\end{split}
\end{equation}

The width of a given $1:\chi$ resonance is therefore:
\begin{equation}
\begin{split}
    \Delta a &= \frac{8 \chi \sqrt{a}}{\sqrt{\mathcal{G}M}} \cdot \sqrt{\frac{\frac{3\mathcal{G} m_N \mathcal{M}_3 \alpha^3}{8a}  X_{\chi}^{-4,1}}{\frac{3n_N\chi}{[L]}}}  = \sqrt{\frac{8m_n}{M_\odot} \underbrace{\mathcal{M}_3}_{\approx 1}} \cdot \sqrt{X_{\chi}^{-4,1}  \alpha^3  \cdot \underbrace{\frac{\chi [L]}{n_N}}_{a^2}}  = \frac{a}{\chi}\sqrt{\frac{8m_n }{M_\odot} X_{\chi}^{-4,1}}
\end{split}
\end{equation}

\section{Derivation of resonance width for arbitrary $m:j$ resonances} \label{sec:mj_res}
For completeness, we give again the disturbing function for the co-planar three-body system with the Sun, Neptune, and an SDO (massless test particle):

\begin{equation}
\begin{split}
    \mathcal{R} = \frac{\mathcal{G} m_N}{a} \sum_{l=2}^\infty \sum_{m=m_{min}}^l \sum_{j'=-\infty}^\infty \sum_{j=-\infty}^\infty & \zeta_m c_{lm}^2 \mathcal{M}_l \times \alpha^l X_{j'}^{l,m}(e_N) X_{j}^{-(l+1),m}(e) \times \\& \cos(j'M_N - jM + m(\omega_N - \omega)).
\end{split}
\end{equation}



For the generic $m:j$ case, we can still make certain simplifications relevant to the circular restricted three-body problem. We start by addressing only the innermost sum, by choosing a order $\ell$, which fixes the order of the expansion. For example, $\ell=2$ is the quadrupole expansion, $\ell=3$ is the octupole expansion and so on. Then, we fix $m=\ell$, which sets the resonant index, such that the quadrupole expansion for $\ell=m=2$ gives a $2:j$ chain of resonances. Choosing an $m<\ell$ begins to include integer multiples of the resonance, essentially expanding the perturbative terms associated with the resonance at that location out to higher order, which we can neglect.  With $\ell, m$ fixed, the disturbing function becomes:

\begin{equation}
\begin{split}
    \mathcal{R}_{\ell,m} &= \frac{\mathcal{G} m_N \mathcal{M}_\ell \alpha^\ell c_{\ell m}^2 }{a} \sum_{j=-\infty}^\infty  X_{j}^{-(\ell+1),m}(e) \times \cos(jM - m(M_N-\omega)).
\end{split}
\end{equation}

The $m:j$ Hamiltonian for an expansion out to order $\ell$ is given by:

\begin{equation}
\begin{split}
    \mathcal{H}_{m:j} &= -\frac{\mathcal{G}M_\odot}{2a} - \frac{\mathcal{G} m_N \mathcal{M}_\ell \alpha^\ell c_{\ell m}^2}{a} \times \sum_{j=-\infty}^\infty  X_{j}^{-(\ell+1),m}(e) \cos(jM - m(n_Nt-\omega)) + \mathcal{T}, 
\end{split}
\end{equation}
where we have added a conjugate action $\mathcal{T}$ that will be useful for removing time dependence. For all terms except that of the $1:j$ resonance, $m=l$. The $1:j$ resonance first appears at octupole order, where $l=3, m=1$. We can use Delaunay Variables to rewrite the Keplerian term as:
\begin{equation}
    -\frac{\mathcal{G}M_\odot}{2a} = -\frac{1}{2}\left(\frac{\mathcal{G}M_\odot}{L} \right)^2.
\end{equation}
We then Taylor expand the Keplerian term around the resonance location through the nominal action $[L] = \sqrt{\mathcal{G} M_\odot a_N (j/m)^{2/3}}$
\begin{equation}
\begin{split}
    &-\frac{1}{2}\left(\frac{\mathcal{G}M_\odot}{{L}}\right)^2  = -\frac{1}{2}\left(\frac{\mathcal{G}M_\odot}{{[L]}}\right)^2 + \underbrace{\left(\frac{\mathcal{G}M_\odot}{{L}}\right)^2\left(\frac{\delta {L}}{[{L}]}\right) - \frac{3}{2}\left(\frac{\mathcal{G}M_\odot}{{L}}\right)^2\left(\frac{\delta {L}}{[{L}]}\right)^2}_{[n]\left(\delta L-\frac{3}{2}\frac{\delta{L}^2}{{[L]}}\right)}
\end{split}
\end{equation}
where $[n] = n_\text{N} \cdot m/\chi$. Here, $\chi$ fixes the index of the specific $m:\chi$ resonance of interest. We can now define a type-2 generating function:
\begin{equation}
    \mathcal{F}_2 = \underbrace{(\chi l/m - (n_N t - g))}_{\phi}\Phi + \underbrace{(l)}_{\psi}\Psi + \underbrace{(t)}_{\xi} \Xi
\end{equation}
with transformation equations between the old and new actions defined as:
\begin{equation}
    \begin{split}
        &\delta L = \frac{\partial \mathcal{F}_2}{\partial l} = \chi \Phi/m + \Psi\\
        &G = \frac{\partial \mathcal{F}_2}{\partial g} = \Phi\\
        &\mathcal{T} = \frac{\partial \mathcal{F}_2}{\partial t} = \Xi - n_N \Phi.
    \end{split}
\end{equation}
We can now apply these transformations to our Hamiltonian, isolating just one resonance $m: \chi$ instead of the series of terms:
\begin{equation}
    \begin{split}
        \mathcal{H}_\chi &= - \frac{\mathcal{G} m_N \mathcal{M}_\ell \alpha^\ell c_{\ell m}^2}{a}  X_{\chi}^{-(\ell+1),m} \cos(m\phi) + \Xi - n_N \Phi + \\
        &[n]\left(\chi \Phi/m + \Psi -\frac{3}{2}\frac{(\chi \Phi/m + \Psi)^2}{{[L]}}\right) -\frac{1}{2}\left(\frac{\mathcal{G}M_\odot}{{[L]}}\right)^2
    \end{split}
\end{equation}
Simplifying and removing constant terms (those without $\Phi$ or conjugate angle $\phi$):
\begin{equation}
    \begin{split}
        \mathcal{H}_\chi &= - \frac{\mathcal{G} m_N \mathcal{M}_\ell \alpha^\ell c_{\ell m}^2}{a}  X_{\chi}^{-(\ell+1),m} \cos(m\phi) - \frac{3n_N \chi}{2m[L]}\Phi^2 - \frac{3n_N}{[L]}\Psi\Phi
    \end{split}
\end{equation}
We can further simplify this by completing the square using the linear action term:
\begin{equation}
    \begin{split}
        \mathcal{H}_\chi &=  \underbrace{\frac{\mathcal{G} m_N \mathcal{M}_\ell \alpha^\ell c_{\ell m}^2}{a}  X_{\chi}^{-(\ell+1),m}}_{\gamma} \cos(m\phi)   - \underbrace{\frac{3n_N\chi}{m[L]}}_{\beta} \cdot \frac{1}{2} \underbrace{\left(\Phi + \frac{m\Psi}{\chi}\right)^2}_{\Tilde{\Phi}^2}  = \gamma \cos{(m\phi)} - \frac{\beta}{2} \Tilde{\Phi}^2
    \end{split}
\end{equation}
where $\Tilde{\Phi}$ is the new action conjugate to the angle $\phi$. We can now calculate the resonance width. For a mathematical pendulum, we have $\Delta \Tilde{\Phi} = 4 \sqrt{\gamma/\beta}$. Using our transformation equations for $\delta L$, $\Delta \delta L = \chi/m \cdot \Delta \Tilde{\Phi}$. However, we want the width, $\Delta a$, which we can find from $\Delta \delta L = \Delta (L-[L]) = \Delta L$, where $[L]$ is constant, such that $\Delta[L]= 0$. Therefore, 
\begin{equation}
\begin{split}
    \Delta \delta L &= \Delta(\sqrt{\mathcal{G}M_\odot a}) = \frac{\Delta a}{\sqrt{a}}\frac{\sqrt{\mathcal{G}M_\odot}}{2}\\
    \Delta a &= \frac{2\sqrt{a} \Delta \delta L}{\sqrt{\mathcal{G}M_\odot}}.
\end{split}
\end{equation}
Combining with $\Delta \delta L = \Delta \Tilde{\Phi} \chi/m $ and $\Delta \Tilde{\Phi} = 4 \sqrt{\gamma/\beta}$, we get:
\begin{equation}
\begin{split}
    \Delta a = \frac{8a^{1/2} \chi}{m\sqrt{\mathcal{G}M_\odot}} \sqrt{\gamma/\beta} &= \frac{8a^{1/2} \chi }{m\sqrt{\mathcal{G}M_\odot}} \sqrt{\frac{\mathcal{G} m_N \mathcal{M}_\ell \alpha^\ell c_{\ell m}^2  X_{\chi}^{-(\ell+1),m}/a}{3n_N\chi/(m[L])}} \\
    &= \frac{8}{\sqrt{3}} \cdot \sqrt{\frac{ \mathcal{M}_\ell \alpha^\ell c_{\ell m}^2 X_{\chi}^{-(\ell+1),m} \chi [L]}{m \cdot n_N} \cdot \frac{m_N}{M_\odot}}  \\ 
    &= \frac{8}{\sqrt{3}} \cdot \sqrt{\frac{m_N}{M_\odot}} \cdot \sqrt{\mathcal{M}_\ell \alpha^\ell c_{\ell m}^2 X_{\chi}^{-(\ell+1),m}} \cdot \sqrt{\frac{ \chi [L]}{m \cdot n_N}} \\
    &= \frac{8a}{\sqrt{3}} \cdot \sqrt{\frac{m_N}{M_\odot}} \cdot \sqrt{\alpha^\ell \mathcal{M}_\ell c_{\ell m}^2 X_{\chi}^{-(\ell+1),m}},
\end{split}
\end{equation}
the expression for resonance width in our paradigm for an arbitrary $m:j$ resonance.
Following \cite{Chirikov}, we can compare the resonance width to that of adjacent resonances (s.t. $\delta a \approx (2a_N/3m)(m/\chi)^{1/3} = (2a_N/3m) \times \alpha^{1/2}$):
\begin{equation}
    \frac{\Delta a}{\delta a} = 4m \cdot \sqrt{\frac{m_N}{M_\odot}} \cdot\sqrt{3\alpha^{\ell-3} \mathcal{M}_\ell c_{\ell m}^2 X_{\chi}^{-(\ell+1),m}}.
\end{equation}
This expression only holds for adjacent resonances of the same order, it does not apply for the $2:10$ and $4:21$ resonance overlap, for example. A more generalized expression can be written down:
\begin{equation}
    \frac{\Delta (a,a')}{\delta (a,a')} = \frac{8\left( \sqrt{\alpha^\ell \mathcal{M}_\ell c_{\ell m}^2 X_{\chi}^{-(\ell+1),m}} +  \sqrt{\alpha^{\ell'} \mathcal{M}_{\ell'} c_{\ell'm'}^2 X_{\chi'}^{-(\ell'+1),m'}}\right)}
    {\sqrt{3}\alpha\left(\left(\chi/m\right)^{2/3} - \left(\chi'/m'\right)^{2/3} \right)}.
\end{equation}

\section{Simulations}
\label{sec:simulations}
We carried out the simulations used for comparison to our analytic model using the \texttt{REBOUND} software package \citep{Rein2012A&A, Rein2019MNRAS}, utilizing the \texttt{TRACE} integration algorithm \citep{Lu2024MNRAS}. We initialize our simulations with: the sun. Neptune with $m=5.15\cdot10^{-5}M_\odot$, $a=30$au, and zero eccentricity, inclination, and mean anomaly, and a test particle SDO on a grid of $a/q$ values with random mean anomaly and zero inclination. We use a timestep corresponding of $\delta t = P_{\text{Neptune}}/(2\pi\times113)$, and integrate for $\Delta t= 2\pi\times10^6$ code units. 

These simulations were then used to derive the MEGNO chaos indicatior (Y) (mean exponential growth of nearby orbits; \citealt{Cincotta2000A&AS}). The MEGNO criterion converges to 2 for stable quasi-periodic motion, and increases with time for chaotic motion. Note here that chaotic motion with $Y>2$ does not directly correspond to an orbit that is unstable on some given timescale. While larger of values of MEGNO will correspond to more chaotic orbits which are likelier to lead to ejection of the SDO, the MEGNO criterion or Lyapunov time are distinct from the diffusion timescale and cannot be used to directly estimate the time until an SDO is ejected from the Solar system.

\section{Approximations to Stability Boundaries}
\label{sec:appxfits}
In order to facilitate the use of our theoretical results, we provide approximate functional fits to the curves shown in \autoref{fig:diagram}. The boundary for the fully chaotic layer wherein objects pass unimpeded through $1:n$ resonances can be approximated with a power-law:
\begin{equation}
    q= 13.37 + 0.55 \times a^{0.64}
\end{equation}
The second curve (red in \autoref{fig:diagram}) which sets the boundary for orbits which have their diffusion slowed by interacting with $1:n$ resonances, can be approximated with a logarithmic function:
\begin{equation}
    q=-63.1+19.4\times\log{(a)}
\end{equation}
Finally, the fine comb-like boundary shown in blue in \autoref{fig:diagram} is approximated by a power-law:
\begin{equation}
    q= 10.28 + 6.01 \times a^{0.34}.
\end{equation}

\end{document}